\documentclass[nofootinbib]{revtex4}

\usepackage{amsmath,amssymb,graphicx}

\newcommand{\tbf}{\textbf}
\renewcommand{\leq}{\leqslant}

\begin{document}

\bibliographystyle{/home/speysson/lib/tex/tex-inputs/revtex4/apsrev}

\title{Finite temperature structure factor in the Haldane-Shastry spin chain}
\author{St\'ephane Peysson}
\affiliation{Institute for Theoretical Phyics, University of Amsterdam, 
Vlackenierstraat 65, 1018 XE Amsterdam, The Netherlands}

\begin{abstract}
The Haldane-Shastry spin chain can be mapped to the
infinite coupling limit of the SU(2) spin Calogero-Sutherland model. We use the
$\mathfrak{gl}_2$ Jack polynomials'
technology to compute the form factors of the spin operator on the multi-spinon 
spectrum. The spin structure factor is obtained through a form factor expansion.
The expansion is proven to converge in the small momentum limit. Numerics based 
on two- and four-spinons contributions give an approximate result for the
infinite temperature static and dynamic spin structure factor.
\end{abstract}

\maketitle

\section{Introduction}

Low-dimensional systems constitute fertile breeding grounds for exotic types of
physical excitations.
Fractionalization of quantum numbers like charge and spin is known to take place
respectively in one-dimensional interacting electron liquids and spin chains: in
cases such as these, one must forget about weakly coupled particles, and instead
adopt a whole new starting point for the description of the strongly coupled
physics.
Obviously, the identification and proper description of this new starting point
is often a very involved and risk-prone process.

In this respect, quasi-one-dimensional spin systems have provided one of the
sturdiest arenas.
Experimental realizations of systems with fractionalized excitations are
numerous and well-documented.
Probably the clearest and best studied signature comes from neutron scattering
experiments on effectively one-dimensional antiferromagnetic spin-1/2 chains
\cite{TennantEXP}.
The excitations seen are not the naively expected spin-1 spin waves, but rather
gapless ``spinons'', which one could losely present as spin-1/2 spin waves.
Among many remarkable properties of these excitations are their fractional
statistics, intermediate between fermions and bosons, making such a system
markedly different from one obeying convential rules.

On the theoretical side, strongly-coupled systems like spin chains have in the
last few decades presented extreme, if not seemingly insurmountable
difficulties.
The simplest way of explaining this fact might be to say that quantum
fluctuations are very strong in one-d, and cannot be tamed by perturbative
approaches.
Instead, excitations are strongly nonlinear, and one is faced with the seemingly
impossible challenge of either providing an exact solution or risking to miss
out completely on the correct physics.

The quantity of interest to experimentalists (thinking about neutron scattering
experiments) is the dynamical spin structure factor (DSSF).
For a chain of $N$ spins at sites $R_i$, this is defined as
\begin{equation}
S^{\alpha \beta} ({\bf q},\omega; T) = \frac{1}{2\pi N} \sum_{i,j} e^{i {\bf q}
\cdot ({\bf R}_j - {\bf R}_i)} \int_{-\infty}^{\infty} e^{i \omega t} \langle
S^{\alpha}_i (0) S^{\beta}_j (t) \rangle_T
\label{structurefactor}
\end{equation}
where the angular brackets denote a thermal average.

The model of choice for the description of the spin dynamics depends of course
on the specifics of the experimental setup one wishes to describe.
The XXZ Heisenberg model \cite{HeisenbergZP49} often fits the bill remarkably
well, at least for very low energies.
The Bethe Ansatz method \cite{BetheZP71} could provide most of its thermodynamic 
properties, but little about its dynamics.
Approximate methods have thus been used to address the computation of the DSSF.
The M\"uller Ansatz \cite{MuellerPRB24} is the best conjecture for the zero
temperature structure factor based on exact results and numerics.
Finite temperature low-energy features were obtained by Schultz using
bosonization \cite{SchultzPRB34}.

To go beyond the field theory limit requires tackling the nonlinear nature of
the original model.
Considerable insight in this direction was provided by the Algebraic Bethe
Ansatz method \cite{JimboBOOK} and the quantum inverse scattering theory 
\cite{KorepinBOOK}. Bougourzi \emph{et al.} used results from the algebraic 
analysis to compute the exact two-spinon contribution to the DSSF of the 
one-dimensional Heisenberg model 
\cite{BougourziPRB54,KarbachPRB55,BougourziPRB57}. More recently, Maillet
\emph{et al} proved multiple integral representations of elementary blocks of
the correlation functions \cite{KitanineNPB641}. Nevertheless, all these 
approaches restrict to zero temperature and no exact thermodynamic limit is 
known. The computation of the DSSF at finite temperature requires a lot of 
further efforts.

However, one of the properties of the Heisenberg model is that the spinons,
though deconfined, still suffer from a residual interaction.
The spinons are thus not truly free excitations obeying fractional statistics.
There exists on the other hand a very convenient alternative approach based on
the Haldane-Shastry model, whose Hamiltonian is \cite{HaldanePRL60,ShastryPRL60}
\begin{equation}
H_{HS} = J \sum_{i < j} \frac{1}{[d(i-j)]^2} \textbf{S}_i \cdot \textbf{S}_j
\label{HS}
\end{equation}
where $d(i) = \frac{N}{\pi} \sin{\frac{\pi i}{N}}$.
It is in the same universality class as the Heisenberg model, and the
long-distance (decaying as $1/R^2$ for large distances) interaction in fact
simplifies things considerably: the spinons form an ideal gas
\cite{HaldanePRL66} of particles obeying fractional exclusion statistics
\cite{HaldanePRL67}.
The DSSF at $T=0$ can in fact be calculated exactly, and is given in the
thermodynamic limit by \cite{HaldanePRL71}
\begin{align}
S^{\alpha \beta}_{HS} ({\bf q}, \omega) &= \frac{\delta_{\alpha \beta}}{2}
\frac{\Theta(\omega_2 (q_{\parallel}) - \omega) \Theta(\omega - \omega_{1-}
(q_{\parallel})) \Theta(\omega - \omega_{1+} (q_{\parallel}))}{\sqrt{(\omega -
\omega_{1-}(q_{\parallel}))(\omega - \omega_{1+}(q_{\parallel}))}}, \nonumber
\\ \omega_{1-} (q_{\parallel}) &= \frac{J}{2} q_{\parallel} (\pi-
q_{\parallel}),
\quad \omega_{1+} (q_{\parallel}) = \frac{J}{2} (q_{\parallel} -
\pi)(2\pi - q_{\parallel}), \quad \omega_2 (q_{\parallel}) = \frac{J}{4}
q_{\parallel} (2\pi - q_{\parallel}).
\end{align}
This $T=0$ formula is made up only of contributions from the two-spinon channel:
all higher channels have vanishing contributions in the zero-temperature limit.

One of the very nice features of the Haldane-Shastry model is that its dynamics
turn out to be much more easily tractable than those of the Heisenberg chain.
The Haldane-Shastry model is but the first representative in a wider class of
solvable models dubbed the SU(N) Haldane-Shastry chains.
These are in turn obtainable as a particular limit of spin Calogero-Sutherland
models, for which an impressive number of exact results are known in the
mathematical literature.
In particular, there exists a Yangian symmetry leading to the identification of
a set of eigenvectors constructed from $\mathfrak{gl}_N$ Jack polynomials
\cite{UglovCMP191}.
In short, the whole set of expectation values and transition matrix elements one
might want to calculate in the Haldane-Shastry model turn out to have a
correspondence in terms of Jack polynomials.
More details on this will be provided in the bulk of the paper.

Thus, this opens the way to the computation of the DSSF (\ref{structurefactor})
at nonzero temperatures for the Haldane-Shastry model.
Our strategy will be to make use of the technology contained in
\cite{UglovCMP191} to compute the form factors involved in the spin-spin
correlation function needed for the DSSF.
This approach has already been used for $T=0$ dynamical properties of the 
Haldane-Shastry spin chain \cite{YamamotoPRL84,YamamotoJPSJ69} and related 
models (spin Calogero-Sutherland model \cite{YamamotoJPA32}, supersymmetric 
$t-J$ model \cite{ArikawaJPSJ68,ArikawaPRL86}).
The present work is the first to address finite temperature dynamics.

However for a generic quantum field theory divergences appear when developping a
correlation function on the Hilbert space. In the context of integrable field
theories (ITF), it has been proposed that it could be rewritten as a sum free of
divergences. The resulting formula, called a 'form factor expansion' (FFE), can
be evaluated using the scattering data of the ITF. There is still on ongoing
discussion about how precisely the FFE can be implemented in ITFs
\cite{LeClairNPB552,DelfinoJPA34,MussardoJPA34,KonikXXX,SaleurNPB567,Castro-AlvaredoNPB636}.
For the case of conformal field theory (CFT), there has been a similar but 
independent proposal for writing finite-temperature correlators in a FFE 
\cite{vanElburgJPA33}. It is based on the fractional statistics of
the quasiparticles building the Hilbert space.
Evidence was put forward that it converges quickly to the
exact (known) result in terms of the number of excited 
quasi-particles \cite{vanElburgJPA33,PeyssonJPA35}. We will prove in the paper
that the latter approach applies in the case of the Haldane-Shastry spin chain.

Our paper is organized as follows.  First, we recall all the necessary
aspects of the computation of form factors for the Haldane-Shastry
model using Jack polynomials.  We then set out to calculate the form
factors themselves, in increasing complexity of spinon channels. The form factor
expansion is then introduced and proved.
We finally put the results together to provide an expression for the
finite-temperature static and dynamical spin structure factors.  Discussions and
conclusions are amassed at the end.

\section{Jack polynomial technology for the spin Calogero-Sutherland model}
Using the freezing trick \cite{PolychronakosPRL70} the SU(2) Haldane-Shastry
model is obtained by the strong coupling limit of the SU(2) spin
Calogero-Sutherland model \cite{HaPRB46}. The latter describes $N$ particles
with coordinates $\{x_i,i=1 \ldots N\}$ moving on a circle of length $N$ with
the Hamiltonian
\begin{equation}
H_{spinCS} = -\frac{1}{2} \sum_{i=1}^N \frac{\partial^2}{\partial x_i^2} + 
\frac{\pi^2}{2N^2}
\sum_{i \leq j} \frac{\beta 
(\beta+P_{ij})}{\sin^2\left(\frac{\pi}{N}(x_i-x_j)\right)}
\end{equation}
where $P_{ij}$ is the SU(2) exchange between particles $i$ and $j$.
Within the freezing trick, the interaction parameter $\beta$ is taken to
infinity. The particles are therefore pinned at $x_i=i$ and interact \emph{via} 
the SU(2) spin exchange.

The spin Calogero-Sutherland model is more tractable than the Haldane-Shastry 
model since it is continuous and all his eigenfunctions have been explicitely
constructed.
We recall in the following the Uglov construction \cite{UglovCMP191} in terms of
$\mathfrak{gl}_2$ Jack polynomials. The mathematical technology they
provide to compute transition matrix elements is then introduced.

\subsection{Yangian Gelfand-Zetlin basis}
We consider the case $N$ even and $N/2$ odd, so that the ground state is unique.
Uglov determined the so-called Yangian Gelfand-Zetlin basis of this model,
orthogonal through the Yangian action. They are labelled by the 
strictly-decreasing
sequences $k=\{k_i,i=1\ldots N\}$, $k^0=\{N/2+2-i,i=1\ldots N\}$ corresponding
to the ground state. $k$ contains information on both momentum and
spin of the excitation. Writing $k_i=2\overline{k_i}+\underline{k_i}$,
$\overline{k_i}\in \boldsymbol{Z}$ represents momentum and $\underline{k_i} \in
\{1,2\}$ color\footnote{We use here a notation different than Uglov. In fact,
this corresponds to the dual representation, called $*$ in his paper. Results
are not altered, as one has to sum both representations to obtain physical
quantities. Our choice for this representation is mostly practical, leading to
states with positive momentum when $\sigma_i$ is positive.}. More precisely,
the momentum, the energy and spin of a state described by $k$ are
\begin{align}
P_k &= \frac{2\pi}{N} \sum_i \overline{k_i} \\
E_k &= \frac{2\pi^2}{N^2} \sum_i [\overline{k_i} + \beta(N+1-2i)/2]^2 \\
S_k &= \frac{1}{2} \sum_i [\delta_{\underline{k_i},2} -
\delta_{\underline{k_i},1}]
\end{align}

For physical applications it is more convenient to work with excitations over 
the ground state.
One identifies the state $k$ with the pair
\begin{equation}
k \equiv \left( \lambda=(k_i-k_N+i-N, \, i=1 \ldots N-1), \, r= k_N-k_N^0
\right).
\end{equation}
It consists of a zero mode $r$ and a partition $\lambda$ (non-increasing 
sequence of positive integers) of length $N-1$. The Hilbert space is spanned by 
all possible pairs. We refer the reader to \cite{UglovCMP191} for the 
expressions of the physical properties in terms of $(\lambda,r)$. They will be 
specified in the next section for the specific case of the Haldane-Shastry spin 
chain.

\subsection{Uglov's isomorphism}
Uglov determined an isomorphism $\Omega$ between the Yangian
Gelfand-Zetlin basis and the $\mathfrak{gl}_2$ Jack Polynomials defined through
\begin{equation}
\Omega(k) = (x_1 \ldots x_n)^r P^{(2\beta+1,2)}_\lambda(\{x_i\})
\end{equation}
where the Jack polynomial $P_\lambda^{(\gamma,2)}$ is the limit $q=-p$,
$t=-p^\gamma$, $p \rightarrow 1$ of the Macdonald polynomial $P_\lambda(q,t)$.
Then one has
\begin{equation}
(k,l)_{(\beta,2)} = \langle \Omega(k),\Omega(l) \rangle_{(\beta,2)}.
\end{equation}
$(\, . \, , \, .\, )_{(\beta,2)}$ is the Yangian scalar product (see 
\cite{UglovCMP191} for a definition)
, and $\langle\, . \, , \, .\, \rangle_{\beta,2}$ is the following scalar
product in the space of symmetric Laurent polynomials
\begin{equation}
\langle f(\{x_i\}), g(\{x_i\}) \rangle_{\beta,2} = \frac{1}{N!} \prod_{j=1}^{N}
\int \frac{dx_j}{2i\pi x_j} \overline{f(\{x_i\})} \left[ \prod_{1\leq k \neq l
\leq N} (1-x_k^2 x_l^{-2})^\beta (1-x_k x_l^{-1}) \right] g(\{x_i\}).
\end{equation}

The physical quantity we study in this work is the action of the
spin operator. Using Uglov's isomorphism, it is given by (we dropped the scalar 
product indices for convenience)
\begin{equation}
( \lambda,r | s^\pm | \mu,r' ) = \frac{1}{N} \sum_{s\in \boldsymbol{Z}} \langle
\lambda,r | p_{2s+1} | \mu,r' \rangle \delta(S_\lambda-S_\mu=\pm 1),
\end{equation}
where $p_m=\sum_{i=1}^{N} x_i^m$ is the power sum symmetric function.
One can identify $s^\pm$ with $(s^+ + s^-)/2$ and
replace the $\delta$ by $1/2\delta(|S_\lambda-S_\mu|=1)$.
In this paper we only study form factors which satisfy this selction rule.
By symmetry
between the contributions of $p_m$ and
$p_{-m}$ (or equivalently the duality relation (see \cite{UglovCMP191}), one may
finally only consider the form factors
\begin{equation}
( \lambda,r | s^\pm | \mu,r' ) \equiv \frac{1}{N} \sum_{s\in \boldsymbol{Z}^+}
\langle \lambda | p_{2s+1} | \mu \rangle \delta_{r,r'}
\delta(|S_\lambda-S_\mu|=1).
\label{eq:ff}
\end{equation}

\subsection{$\mathfrak{gl}_2$ Jack technology}
To compute transition matrix elements such as (\ref{eq:ff}),
we need some results from the mathematical
literature on Macdonald polynomials \cite{Macdonald}.

\begin{figure}[h]
\begin{center}
\begin{picture}(120,140)
\put(0,140){\line(1,0){120}}
\put(0,120){\line(1,0){120}}
\put(0,100){\line(1,0){80}}
\put(0,80){\line(1,0){80}}
\put(0,60){\line(1,0){60}}
\put(0,40){\line(1,0){20}}
\put(0,20){\line(1,0){20}}
\put(0,0){\line(1,0){20}}
\put(0,0){\line(0,1){140}}
\put(20,0){\line(0,1){140}}
\put(40,140){\line(0,-1){80}}
\put(60,140){\line(0,-1){80}}
\put(80,140){\line(0,-1){60}}
\put(100,140){\line(0,-1){20}}
\put(120,140){\line(0,-1){20}}
\put(28,88){$s$}
\put(-8,88){$i$}
\put(28,144){$j$}
\put(84,88){$\lambda_i$}
\put(28,50){$\lambda_j'$}
\end{picture}
\end{center}
\caption{Partition  $\lambda = (6,4,4,3,1,1,1).$}
\label{fig:partition}
\end{figure}
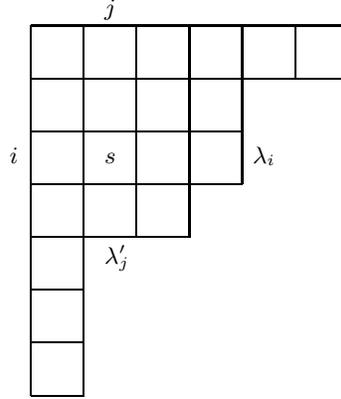

We consider a generic partition as on Fig. \ref{fig:partition}.
It is a tableau made of cases $s$ labeled by their row number $i$ and their 
column number $j$. The length (the number of cases) of the row $i$ is 
$\lambda_i$ and the length of the column $j$ is $\lambda'_j$.
One then defines
\begin{description}
\item[Definitions for partitions]:\\
Length: $l(\lambda)=\text{max}(\lambda_j')$;\\
Cardinal: $|\lambda|=\sum\lambda_i$ and we note $\lambda\vdash |\lambda|$;\\
Arm-length: $a(s)=\lambda_i-j$;\\
Leg-length: $l(s)=\lambda_j'-i$;\\
Arm-colength: $a'(s)=j-1$;\\
Leg-colength: $l'(s)=i-1$;\\
Content: $c(s)=a'(s)-l'(s)$;\\
Hook-length: $h(s)=a(s)+l(s)+1$;\\
$C(\lambda)=\{s\in\lambda | c(s)\equiv 0 \text{ mod }2\}$;\\
$H(\lambda)=\{s\in\lambda | h(s)\equiv 0 \text{ mod }2\}$.
\end{description}

The quantities of interest in the present work are the norm of a given state, 
the expansion of power sums on Jack polynomials (thus on physical states), and 
finally the Pieri formula. The latter gives the development on Jack 
polynomials of the product of a Jack polynomial and an elementary function 
$e_r=P^{(\gamma,2)}_{1^r}$. This is the only formula available to compute form 
factors in the space of symmetric polynomials.

They read

\begin{description}
\item[Norm]
\begin{equation}
\langle P^{(\gamma,2)}_\lambda | P^{(\gamma,2)}_\lambda \rangle = \langle 1 |
1\rangle \prod_{C(\lambda)}
\frac{a'(s)+\gamma(N-l'(s))}{a'(s)+1+\gamma(N-l'(s)-1)}
\prod_{H(\lambda)}\frac{a(s)+1+\gamma l(s)}{a(s)+\gamma(l(s)+1)}
\end{equation}
\item[Expansion of power sums]
\begin{align}
\label{eq:pm-normal}
p_{2s+1} &= \sum_{\lambda\vdash 2s+1} \chi_\lambda P^{(\gamma,2)}_\lambda,\\
\chi_\lambda &= (-)^{n(\lambda)} \frac{\prod_{C(\lambda)\backslash(1,1)}
(a'(s)-\gamma l'(s)}{\prod_{H(\lambda)} a(s)+1+\gamma l(s)} \text{   for
}|C(\lambda)|=|H(\lambda)|+1
\end{align}
\item[Pieri formula]
\begin{align}
\label{eq:pieri}
P^{(\gamma,2)}_\mu e_r &= \sum_\lambda \psi'_{\lambda/\mu} \,
P^{(\gamma,2)}_\lambda,\\
\psi'_{\lambda/\mu} &= \prod_{C_{\lambda/\mu}\backslash R_{\lambda/\mu}}
\frac{b_\lambda(s)}{b_\mu(s)},\\
b_\lambda &= \frac{a(s)+\gamma(l(s)+1)}{a(s)+1+\gamma l(s)}\text{  if }s\in
H(\lambda)\text{, 1 otherwise},
\end{align}
with $\lambda-\mu$ being a vertical $r$-strip (at maximum 1 box per row, for a
total of $r$), $C_{\lambda/\mu}$ (resp. $R_{\lambda/\mu}$) being the union of
columns (resp. rows) that intersect $\lambda-\mu$.
\end{description}

\section{Transition matrix elements in the Haldane-Shastry model}

We now specify this mathematical background to the case of the Haldane-Shastry
spin chain. The $\beta \rightarrow \infty$ limit simplifies a great deal the
algebra. The eigenstates described above can be interpreted as multi-spinon 
states. We give their physical properties in the following. Then a general 
expression for the matrix elements of the spin operator is presented along with 
closed analytical result in the case of few-spinon states.

\subsection{Spinon interpretation}

In the Haldane-Shatry framework, all physical quantities rewrite in terms of 
sums over the columns $\lambda'_j$ of the partition $\lambda$:
\begin{align}
P_\sigma &= \pi r + \pi \sum_j \frac{2 w_j}{N} \\
E_\sigma &= \pi^2 \left[ (1 -(-)^j/N) \frac{2 w_j}{N} - \left(\frac{2
w_j}{N}\right)^2  + (1-(-)^r)/2N \right]\\
S_\sigma &= \sum_j [\lambda'_j - 2 w_j] \\
w_j &= \left\lbrace \begin{array}{l l} \lfloor \frac{\lambda'_j}{2} \rfloor &
\text{for } j+r \text{ even} \\ \lceil \frac{\lambda'_j}{2} \rceil & \text{for }
j+r \text{ odd} \end{array} \right.
\end{align}
At the
thermodynamic limit (which only is of interest), distinctions between even and
odd disappear for the momentum and the energy. Defining $x_j = \pi\lambda'_j /N
\in [0,\pi]$, they are
\begin{align}
P_\sigma &= \pi r + \sum_j x_j \\
E_\sigma &= \sum_j x_j (\pi-x_j)
\end{align}
One recognizes the dispersion relation of spinons.
An excitation can then be described by a zero mode $r$ and a set of particles
called spinon defined by each column of a tableau $\lambda$. We will for the 
sake of simplicity write such a state $\{ m_j = \lambda'_j \}_r$. $r$ can be 
discarded in most of the physical applications.

Now we can express all the Jack polynomial technology in terms of the spinons' 
quantum numbers $m_j$. We will use the short cut notations
\begin{align}
\gamma_{\text{even}} (m) &= \frac{\Gamma(\lfloor m/2 \rfloor +1)
\Gamma(1/2)}{\Gamma(\lfloor m/2\rfloor +1/2)} \simeq
\sqrt{\frac{Nx}{2}}
\\
\gamma_{\text{odd}} (m) &= \frac{m}{2\gamma_{\text{even}} (m)} \simeq
\frac{1}{\pi} \sqrt{\frac{Nx}{2}}
\end{align}

The norm is
\begin{equation}
\label{eq:norm}
N_{\{m_j, \, j=1\ldots n \}} \equiv \frac{\langle P_{\{m_j\}} | P_{\{m_j\}}
\rangle}
{\langle 1 | 1\rangle} = \prod_{i=1}^n \frac{\gamma_{i+1}(N)}
{\gamma_{i+1}(N-m_i)}
\prod_{1 \leq i
\leq j \leq n} \frac{\gamma_{i-j}(m_i-m_j)}{\gamma_{i-j}(m_i - m_{j+1})} \simeq
\prod_i \sqrt{\frac{2\pi}{N E_i}}
\end{equation}

One shows that the power sum operators
decompose into 2-spinon states $(m_1,m_2)$ such that $m_1+m_2=2s+1$ with
\begin{equation}
\label{eq:chi}
\chi_{(m_1,m_2)}=(-)^{m_2} \gamma_0(m_1-m_2).
\end{equation}

Specifying the Pieri formula to $\mu \equiv (m_1,\ldots,m_n)$, $\lambda
\equiv(p_1,\ldots,p_{n+1})$ --- with possibly $p_{n+1}=0$ --- it gives
\begin{equation}
\psi'_{\lambda/\mu} = \prod_{1 \leq i \leq j \leq n}
\frac{\gamma_{i-j}(p_i-p_{j+1}) \gamma_{i-j}(m_i-m_j)}{\gamma_{i-j}(p_i-m_j)
\gamma_{i-j}(m_i-p_{j+1})}
\end{equation}

\subsection{Matrix elements}

Different strategies apply to the evaluation of the transition matrix elements
$\langle \lambda | p_{2s+1} | \mu \rangle$. The first one is to write $\mu$ and
$\lambda$ into elementary functions $e$ by inversion of the Pieri formula
(\ref{eq:pieri}), use the decomposition (\ref{eq:pm-normal}), and apply the
Pieri formula successively on the multi-$e$ state. This solution is quite
unpractical, because inverting the Pieri formula becomes increasingly difficult
with the number of spinons considered. It is trivial for 1 spinon, and leads to
a rather cumbersome expression already for 2 spinons. Still, we can use it to
rewrite the power sums as a sum of 2-$e$ states. The result is interestingly
simple
\begin{equation}
\label{eq:pm-e}
p_{2s+1} =  \frac{1}{2} \sum_{r=0}^s (-)^{s+r+1} (2r+1) e_{s+r+1} e_{s-r}
\end{equation}
To prove it, one uses (\ref{eq:pieri}) on (\ref{eq:pm-e}), it leads to
(\ref{eq:chi}) thanks to the equality
\begin{equation}
1=\frac{1}{2}\sum_{r=0}^s\frac{1}{\gamma_0(r) \gamma_0(s-r)}
\end{equation}

It gives way to a generic strategy: use decomposition (\ref{eq:pm-e}) and use
the Pieri formula twice. The result is (with normalized states)
\begin{equation}
\label{eq:trans}
\langle \lambda | p_{2s+1} | \mu \rangle = \frac{1}{2} \sum_\nu
(-)^{|\lambda|-|\nu|} (|\mu|+|\lambda| - 2 |\nu|) \Psi'_{\lambda \backslash \nu}
\Psi'_{\nu\backslash \mu} \, \sqrt{\frac{N_\lambda}{N_\mu}}
\end{equation}
with $|\lambda|-|\mu|=2s+1$.

We will now evaluate the contribution of several channels. Eq. (\ref{eq:trans})
doesn't give a closed analytic expression for the form factors. Only in a few
cases can it be so reduced. The results may be conjectural, then confirmed
through the following sum rule
\begin{equation}
\label{eq:sum-rule}
\sum_{m>0} \langle \mu | p_m p_{-m} | \mu \rangle \xrightarrow[N \rightarrow
\infty]{} |\mu|
\end{equation}
which can be proved by simple Jack polynomials' algebra.

\begin{description}

\item[0 $\rightarrow$ 2 spinons]

This is the only channel present at zero temperature. It has been conjectured by
Haldane, then proved using the simplectic ensemble \cite{HaldanePRL71}, and
finally
Yamamato \emph{et al.} \cite{YamamotoJPSJ69} obtained it using  Uglov's
technology.

We recall the result and give the thermodynamic limit
\begin{equation}
\langle (m_1,m_2) | p_{m_1+m_2} | 0 \rangle =
\sqrt{\frac{\gamma_0(m_1-m_2)\gamma_1(m_1-m_2)\gamma_0[N]\gamma_1[N]}
{\gamma_1(m_1)\gamma_0(N-m_1)\gamma_0(m_2)\gamma_1(N-m_2)}} \simeq
\sqrt{\frac{\pi(x_1-x_2)}{\sqrt{E(m_1)E(m_2)}}}
\end{equation}

\item[1 $\rightarrow$ 1 spinon]

\begin{equation}
\langle (m') | p_{m'-m} | (m) \rangle =
\sqrt{\frac{\gamma_0(m')\gamma_0(N-m)}{\gamma_0(m)\gamma_0(N-m')}} \simeq \left(
\frac{x' (\pi-x)}{x(\pi-x')} \right)^{1/4}
\end{equation}

\item[2 $\rightarrow$ 2 spinons]

\begin{align}
\langle (m'_1,m'_2) | p_m | (m_1,m_2) \rangle &= \sqrt{\frac{\gamma_0(m'_2)
\gamma_1(m'_1) \gamma_0(N-m_1) \gamma_1(N-m_2)}{\gamma_0(m_2) \gamma_1(m_1)
\gamma_0(N-m'_1) \gamma_1(N-m'_2)}} \nonumber \\
&\times \begin{cases}
\sqrt{\frac{\gamma_0(k)\gamma_1(k')}{\gamma_0(k')\gamma_1(k)}} & \text{if
$m'_1=m_1$}, \\
\sqrt{\frac{\gamma_0(k')\gamma_1(k)}{\gamma_0(k)\gamma_1(k')}} & \text{if
$m'_2=m_2$}, \\
\sqrt{\frac{\gamma_0(k')\gamma_1(k)}{\gamma_0(k)\gamma_1(k')}} \,
G(m,l=m'_2-m_2,k,k') & \text{otherwise}.
\end{cases}
\\
G(m,l,k,k')&=\sum_{i=1}^l (-)^{i(k+l+1)} \frac{\gamma_1(i)}{\gamma_0(i)} \,
\frac{\Gamma(\lceil \frac{m-l+i}{2} \rceil)}{\Gamma(\lfloor \frac{m-l}{2}
\rfloor +1 )} \,
\frac{\Gamma(\lceil\frac{l}{2}\rceil)}{\Gamma(\lfloor\frac{l-i}{2}\rfloor+1)} \,
\frac{\Gamma( \lfloor \frac{k-i}{2} \rfloor
+\frac{1}{2})}{\Gamma(\lfloor\frac{k}{2}\rfloor+\frac{1}{2})} \,
\frac{\Gamma(\lceil\frac{k'}{2}\rceil+\frac{1}{2})}{\Gamma(\lceil\frac{k'+i}{2}\rceil +\frac{1}{2})}
\end{align}
with $m=m'_1+m'_2-m_1-m_2$, $k=m_1-m_2$, $k'=m'_1-m'_2$.
This can be interpreted as an SU(2) generalization of result (28) of
\cite{PeyssonJPA35} for $g=1/2$ quasiparticles.

\item[Higher channels]

It is generally not possible to obtain a closed analytical expression for the
other channels, except in a few cases. Such are, for example, $[(m_1)\rightarrow
(m_1,n,n')]$ and $[(m_1)\rightarrow (n,n',m_1)]$ which equal $[0 \rightarrow
(n,n')]$. Nevertheless, formula (\ref{eq:trans}) can be used to give exact
numerical results. Getting the thermodynamic limit directly is a 
challenging but fruitful work. We leave it as an open question.

\end{description}

\section{Correlation functions at finite temperature}

Before addressing the computation of the DSSF, general considerations on 
finite-temperature correlation functions are needed. For a local operator 
$\mathcal{O}$, it is
\begin{equation}
\label{eq:1point}
\langle \mathcal{O} \rangle_T = \frac{\sum_n \sum_{\lambda_n} \langle
\lambda_n | \mathcal{O} | \lambda_n \rangle \exp (-\beta E_{\lambda_n})}
{\sum_n \sum_{\lambda_n}\exp (-\beta E_{\lambda_n})}
\end{equation}
where $n$ is the number of quasiparticles, and $\lambda_n$ a state
with $n$ quasiparticles.

As recalled in the introduction, divergences appear in the correlators of the 
right-hand side that need to be resummed. For ITFs, LeClair and Mussardo 
proposed such a resummation as a form factor expansion on the basis of the 
asymptotic particle states in the zero-temperature theory \cite{LeClairNPB552}
\begin{equation}
\label{eq:FFE1}
\langle \mathcal{O} \rangle_T = \sum_n \sum_{\lambda_n} \langle
\lambda_n | \mathcal{O} | \lambda_n \rangle_\text{irr}
\prod_{i=1}^{n} \bar{n}_T (E_{\lambda_i})
\end{equation}
The irreducible form factor $\langle
\lambda_n | \mathcal{O} | \lambda_n \rangle_\text{irr}$ is obtained thanks to
the Form Factor Bootstrap (FFB), and $\bar{n}_T (E)$ is the filling factor 
determined by the Thermodynamic Bethe Ansatz (TBA).

In the following, we show that the FFE apply also for the Haldane-Shastry spin 
chain in the thermodynamic limit. First we obtain the thermodynamic properties 
of the spinons, then give the expression of the irreducible form factor.

\subsection{Exclusion statistics}

A form factor expansion similar to (\ref{eq:FFE1}) was proposed for CFTs 
\cite{vanElburgJPA33}. It led to identify a two-body S-matrix for the CFT in the 
thermodynamic limit
\begin{equation}
\boldsymbol{S}=\exp[2i\pi (\boldsymbol{\delta}-\boldsymbol{K}) \Theta(\theta)]
\end{equation}
where $\boldsymbol{K}$ is the exclusion statistics' matrix of the 
quasi-particles of the theory (which play the role of asymptotic states).

Fractional exclusion statistics is a tool introduced by Haldane
\cite{HaldanePRL67} for the analysis of strongly correlated many-body systems.
It is only based on the assumption that the Hilbert space is finite-dimensionnal
and extensive, i.e. particles are excitations of the considered condensed matter
system, so it is a very generic concept.
The statistics are encoded in a matrix $\boldsymbol{K}=(K_{ij})$ correponding to
the reduction of the available Hilbert space for particle of type $i$ by filling
a one-particle state by a particle of type $j$.
This is then a generalization of the Pauli principle.

For spin-1/2 spinons with species $i=\pm$, the statistical matrix is 
\cite{BouwknegtNPB547}
\begin{equation}
\boldsymbol{K}=\left(\begin{array}{cc} \frac{1}{2} & \frac{1}{2} \\ \frac{1}{2}
& \frac{1}{2} \end{array} \right)
\end{equation}
As such, their 1-particule distribution functions generalize the
familiar Fermi-Dirac and Bose-Einstein ones. They are derived from 1-particle
grand canonical partition functions $G_i$ given by the IOW equations 
\cite{IsakovMPLB8,DasnieresdeVeigyPRL72,WuPRL73}
\begin{equation}
\left(\frac{G_i-1}{G_i}\right) \prod_j G_j^{\boldsymbol{K}_{ij}} = z_i
\end{equation}
where $G_i$ depends on the generalized fugacities $z_j=
e^{\beta(\mu_j-\varepsilon)}$. The one-particule distribution functions are
obtained through
\begin{equation}
n_i(\varepsilon) = z_i \frac{\partial}{\partial z_i} \log \prod_j G_j
\end{equation}

In our case, where species are $i=(+,-)$, we obtain
\begin{equation}
n_\pm (\varepsilon)= \frac{z_\pm}{\sqrt{1+\left(\frac{z_+-z_-}{2}\right)^2}}
\,\frac{\pm \left(\frac{z_+-z_-}{2}\right) +
\sqrt{1+\left(\frac{z_+-z_-}{2}\right)^2}}{\left(\frac{z_++z_-}{2}\right) + 
\sqrt{1+\left(\frac{z_+-z_-}{2}\right)^2}} \xrightarrow{\mu^+=\mu^-} 
\frac{1}{\exp(\beta E) +1}
\end{equation}
from which one deduces that at zero magnetic field,
these distributions match the
Fermi-Dirac distribution function.
It means that spinons can be considered as having fermionic statistics.
This is indeed what is done when they are labeled by ordered numbers within
partitions, the spin degree of freedom is hidden. The norm formula
(\ref{eq:norm}) shows
that for a multi-spinon state, the labels have to be strictly decreasing, as for
spinless fermions.

\subsection{Irreducible form factors}

For ITFs, the definition of the irreducible form factors comes \emph{a priori}
from the FFB. Within our framework (based on \cite{vanElburgJPA33}) it is not 
even necessary. Their definition is
\begin{equation}
\label{eq:irr}
\langle \lambda_n | \mathcal{O} | \lambda_n \rangle = \langle
\lambda_n | \mathcal{O} | \lambda_n \rangle_\text{irr}
+ \sum_{\overline{\lambda_n}}
\langle \overline{\lambda_n} | \mathcal{O} | \overline{\lambda_n}
\rangle_\text{irr}
\end{equation}
with $\overline{\lambda_n}$ being a sub-state of $\lambda_n$ (a substate being a 
state where some of the spinons have been taken out).
Nonetheless, the FFB insures that irreducible form factors don't carry
divergences, which we can't prove here. This calls for further understanding of 
form factors in fractional statistics' theories.

Here follows a sketch of the proof that the irreducible form factors 
(\ref{eq:irr}) give the correct FFE (\ref{eq:FFE1}). It only uses the fact that 
the colorless spinons are free fermions, which shows up by the strict ordering 
of their quantum numbers within a multi-spinon state. As such, the proof is
similar to \cite{LeClairNPB482}, but in the discrete case.

We first remark that $\langle 0 | \mathcal{O} |0 \rangle_\text{irr}$ trivially
comes with
the factor 1. Let us consider the factor of
$D(m)=\langle (m) | \mathcal{O} | (m) \rangle_\text{irr}$.
The contribution $C_n$ from $n$ quasi-particles is obtained recursively, 
isolating the spinons whose label match with $m$:
\begin{align}
C_n &= \frac{1}{Z} \sum_{m_1>\ldots >m_n} \sum_i D(m_i) \prod_{i=1}^{n}
\exp(-\beta E_{m_i})\\
&= \frac{1}{Z} \sum_{m_1>\ldots >m_{n-1}} \prod_{i=1}^{n-1} e^{-\beta E_{m_i}}
\sum_m D(m) e^{-\beta E_{m}} (1-\sum_{i=1}^{n-1} \delta_{m,m_i})\\
&= \frac{1}{Z} \left[\sum_{m_1>\ldots >m_{n-1}} \prod_{i=1}^{n-1}
e^{-\beta E_{m_i}} \sum_m D(m) e^{-\beta E_{m}} \right. \nonumber \\ &- \left.
\sum_{m_1>\ldots >m_{n-2}} \prod_{i=1}^{n-2} e^{-\beta E_{m_i}} \sum_m D(m)
e^{-2\beta E_{m}}(1-\sum_{i=1}^{n-2} \delta_{m,m_i}) \right]\\
&= \ldots = \frac{\sum_m D(m) e^{-\beta E_{m}}
\sum_{i=0}^{n-1} (-)^i z_{n-1-i} e^{-i\beta E_{m}}}{\sum_{i=0}^{\infty} z_i}\\
z_i&= \sum_{m_1>\ldots >m_i} \prod_{j=1}^{i} \exp(-\beta E_{m_j}).
\end{align}
Then summing over $n$ gives
\begin{align}
\sum_{n=1}^{\infty} C_n &= \sum_m D(m) e^{-\beta E_{m}}
\frac{\sum_{n=1}^{\infty} \sum_{i=0}^{n-1} (-)^i z_{n-1-i} e^{-i\beta
E_{m}}}{\sum_{i=0}^{\infty} z_i} \\
&=\sum_m D(m) e^{-\beta E_{m}} \frac{\sum_{i=0}^{\infty} (-)^i
e^{-i\beta E_{m}}
\sum_{n=i+1}^{\infty} z_{n-1-i} }{\sum_{i=0}^{\infty}
z_i} \\
&=\sum_m D(m) e^{-\beta E_{m}} \sum_{i=0}^{\infty} (-)^i e^{-i\beta E_{m}}\\
&= \sum_m D(m) \frac{1}{\exp(\beta E_m) +1}
\end{align}

The demonstration follows the same lines for the higher form factors.
Eq. \ref{eq:FFE1} is quite convenient. Indeed, the denominator is suppressed,
so it can be seen as a perturbative series, provided that it's converging. It
also has a clear physical interpretation with quasi-particles and
filling-factors. Now it can be applied to the particular case of the dynamical
spin structure factor (DSSF).

\section{Finite-temperature spin structure factor}

Now having all the methodological ingredients we address the main subjet of our 
paper: the computation of the finite-temperature spin structure factor of the 
Haldane-Shastry spin chain. We first recall the zero-temperature result. It was 
first obtained by Haldane and Zirnbauer \cite{HaldanePRL71} using the 
supermatrix method. It can also be accessed with $\mathfrak{gl}_2$ Jack 
polynomials \cite{YamamotoPRL84,YamamotoJPSJ69}.
Only the $[0\rightarrow 2]$ channel
contributes so it is straightforwardly
\begin{equation}
S_0(q,\omega)=\int_0^\pi dx_1\int_0^\pi dx_2 
\frac{\pi |x_1-x_2|}{\sqrt{E(x_1)E(x_2)}} \,\delta (x_1+x_2-q) \,\delta
\left( E(x_1)+E(x_2)-\omega \right)
\end{equation}
At finite temperature, all channels contribute, one needs to sum them all 
through a FFE. We detail in the following the static structure factor 
$S_T(q)=\int d\omega \, S_T(q,\omega)$ and the DSSF.

\subsection{Static structure factor}

The static spin-spin correlator can be described by a one-point function
\begin{equation}
\label{eq:stat}
S_T(q=2\pi s/N) = \langle \frac{1}{N} p_{-(2s+1)} p_{(2s+1)} \rangle_T
\end{equation}
(this is not the exact one, but equal at the thermodynamic limit)
To compute it, we use the 1-point FFE and the transition matrix elements
obtained before.

Due to the complexity of the form factor and the lack of FFB, it is hardly
impossible to obtain analytical results on the correlation functions.
The only point for which we could obtain the correlator exactly is $q=0$.
We compute it as the thermodynamic limit ($N\rightarrow\infty$) of 
(\ref{eq:stat}) at $s=0$. The FFE is
\begin{align}
\langle p_{-1} p_1 \rangle_T &= \frac{1}{N}\sum_\mu n_F(E_{\mu}) \langle
\mu| p_{-1} p_1|\mu \rangle_{\text{irr}} \\
&=\sum_{n=1}^\infty F_n \\
F_n &= \sum_{m_1>\cdots>m_n} \prod_{i=1}^n n_F(E_{m_i}) \langle
\{m_1,\ldots,m_n\}|p_{-1}p_1|\{m_1,\ldots,m_n\}
\rangle_{\text{irr}}
\end{align}
In the appendix we show that
\begin{equation}
\label{eq:Fn}
F_n=\int_0^\pi dx [n_F(E(x))]^n
\end{equation}
from which we conclude
\begin{align}
S_T(q=0)
&=\sum_{n=1}^\infty \int_0^\pi dx \left(\frac{1}{\exp(\beta E(x)) +1}\right)^n
\\
&= \int_0^\pi dx \, e^{-\beta E(x)}
\end{align}
This result brings strong evidence of the power of the FFE to obtain 
finite-temperature correlation functions. This is the main achievement of the 
paper. Let us remark here that $F_n$ is the contribution of the $n$-spinon 
states in
the FFE. They appear to be of the same order in temperature, meaning that the
FFE is not a low-temperature expansion\footnote{This would have been expected on 
the ground that more thermal energy is needed to excite more particles. This is
not true anymore for massless particles.}. It really is a
perturbative expansion,
in the sense that the different contributions decrease exponentially with the
number of spinons involved.

Nonetheless, $S_T(q=0)$ is a low-energy feature of the theory that can be
obtained 
with a bosonization approach. As in \cite{PeyssonJPA35}, we observe that the FFE 
is mostly powerful in regimes where other simpler methods apply.
To calculate the whole static structure factor, we then
rely on numerics. We work in the infinite-temperature regime to compare the FFE 
with the expected result. In this limit, correlations are only local and we
expect the static structure factor not to depend on the momentum and have the 
value $S_{T=\infty}(q=0)=1$.

\begin{figure}[htb]
\begin{center}
\includegraphics[width=12cm]{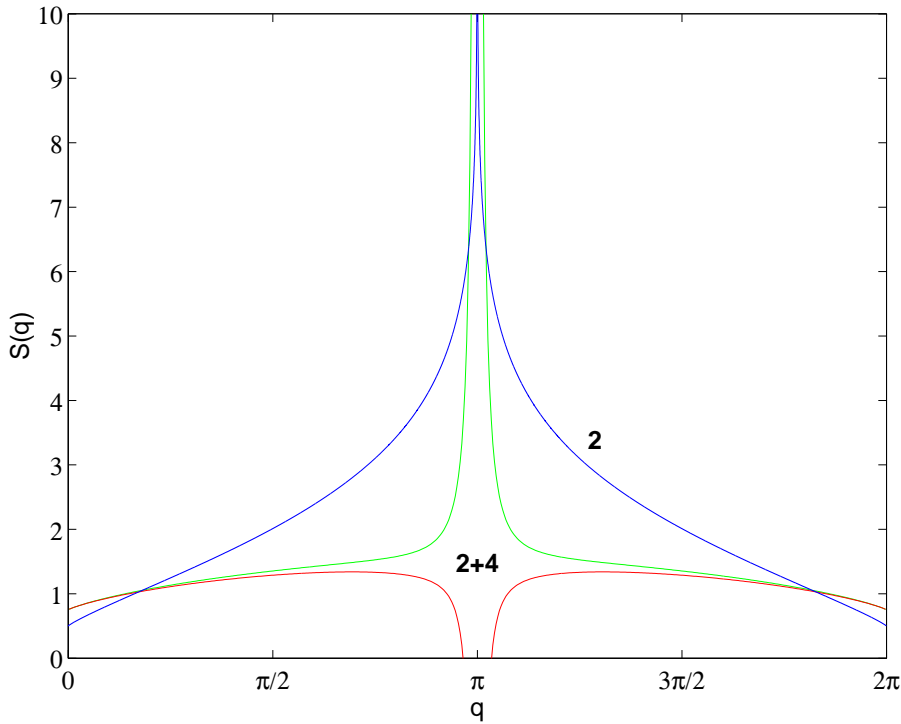}
\end{center}
\caption{Static spin structure factor. \tbf{2} is the contribution from
2-particle form factors with $N=10000$; \tbf{2+4} is the contribution from
2- and 4-particle form factors with $N=500$; the two different curves are
obtained with two labellings of the spinons: for the upper curve, $0<m<N$ (the 
normal one), for the lower curve, $0 \leq m \leq N$; they should match in the 
$N\rightarrow \infty$-limit (up to a $\delta$-peak on $q=\pi$), so we expect the 
thermodynamic limit to lie inbetween the two curves.}
\label{fig:stat}
\end{figure}

To observe the FFE perturbative power, we studied the contribution from 2- and
(2+4)-multispinon states. What we understand here as 2-spinons is the sum of the
contributions of the $[0\leftrightarrow2]$ and $[1\leftrightarrow1]$ channels,
and as 4-spinons the $[1\leftrightarrow3]$ and $[2\leftrightarrow2]$ channels.
Results are gathered on figure \ref{fig:stat}.
At first we observe the convergence to be rather good. This strongly supports 
our approach. Still problems arise in the vicinity of $q=\pi$. Finite-size 
effects starts appearing at the level of 4-spinons contributions, and increase 
with a higher number of spinons. The only way to tackle them is to have an 
enormous computing time. We admit this is the most important weakness of the
approach. This emphasizes the need for a direct thermodynamic limit computation.

\subsection{Dynamic structure factor}
\begin{figure}[htb]
\begin{center}
\includegraphics[width=13cm]{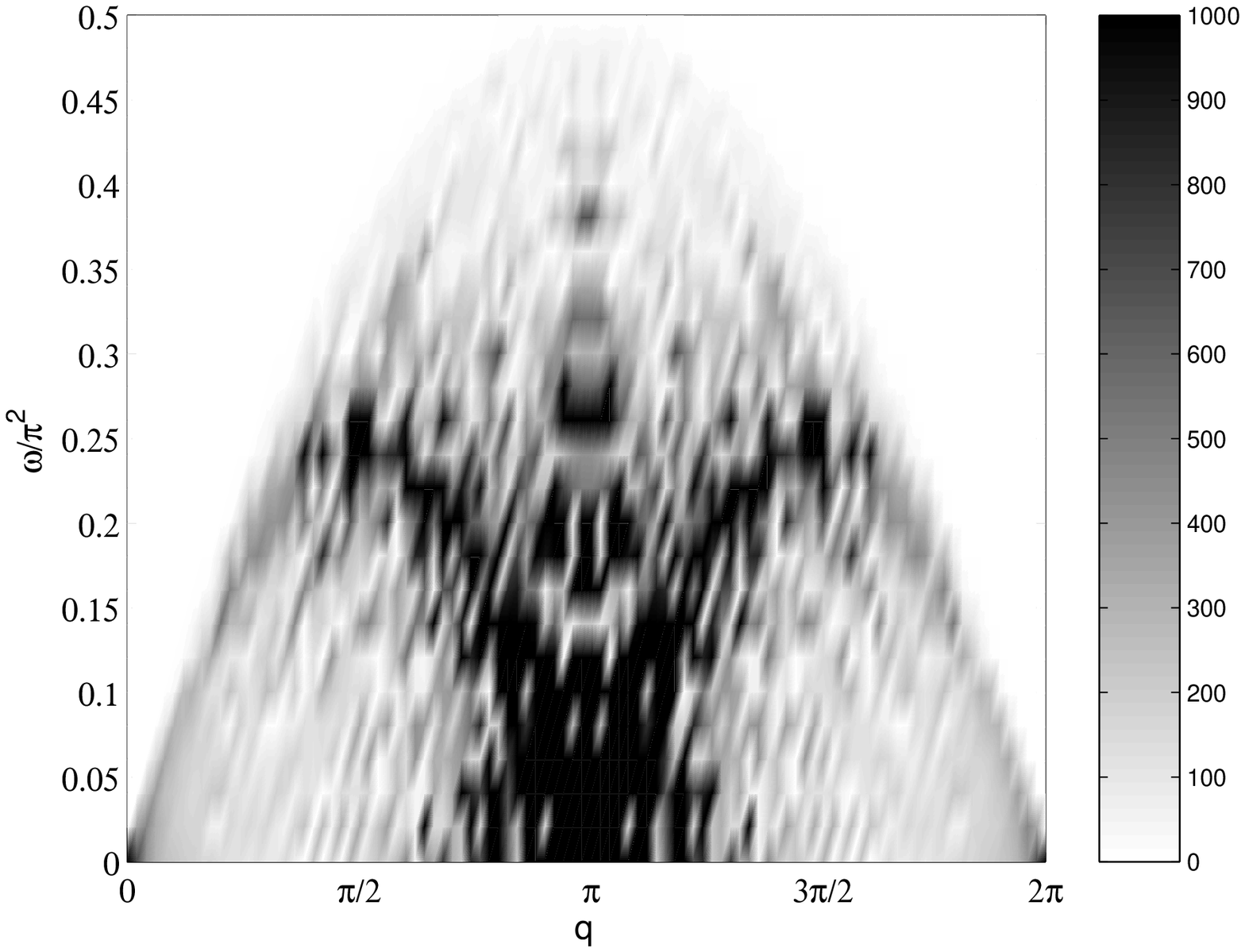}
\end{center}
\caption{Dynamic spin structure factor: the contributions from
2- and 4-particle form factors with $N=500$, in absolute value.}
\label{fig:dyn}
\end{figure}

We show on figure \ref{fig:dyn} the sum of the contributions from 2- and 
4-spinons to the infinite-temperature DSSF.
As in the previous paragraph, finite-size effetcs clearly appear around $q=\pi$.
This makes the comparison to experiments hardly possible. We still note that 
some basic features, like the double arch shape, agree qualitatively. This 
feature, already present at zero temperature,
comes from form factors where the spin operator acts on only one
of the spinon in a multi-spinon excitation, leading to the spinon dispersion
relation.
We leave more comments to the concluding section.

\section{Conclusion}

In this paper we studied the finite-temperature spin-spin correlation function 
of the Haldane-Shastry spin chain. We used the multi-spinon basis obtained as 
the infinite coupling limit of the Yangian basis of the spin Calogero-Sutherland
model. Form factors of the spin operators were computed thanks to the
$\mathfrak{gl}_2$ Jack polynomial technology. These form factors were gathered 
within a form factor expansion of finite-temperature correlation functions to 
give physical quantities directly comparable with the experiments. Though the 
Haldane-Shastry is not a quantitatively appropriate model to describe real spin 
chains such as KCuF$_3$, it should at the qualitative level. As any 
finite-temperature properties for the Heisenberg model are yet beyond reach, 
serious theoretical insights are brought by the Haldane-Shastry model. We recall 
that it is simpler due to the $1/r^2$ interaction responsible for the
thermodynamic freedom of the spinons.

Our main analytical results are: the formal expression of any form factor of the
theory, a closed analytical expression for the simplest of them, the proof of 
the finite-temperature FFE, and the value at zero-momentum of the static 
structure factor. Along with numerical results, they make evidence of the FFE 
technique to approximate finite-temperature correlations. As a drawback, it
demands huge computation power. This calls for further theoretical refinement.

Work is on progress to provide the Haldane-Shastry model with integrability 
features such as the FFB. Within such a description, thermodynamic irreducible 
form factors would be obtained much easier using scattering properties. Problems 
to develop this approach come from the intrinsic discrete nature of the 
Haldane-Shastry spinons. But necessary effort has to be made to gather 
comprehensively ITFs and CFTs in a common framework.
Further understanding is also needed in the link between the Haldane-Shastry
model and the symplectic random matrix ensemble used in \cite{HaldanePRL71}. The
analysis performed at zero temperature could be extented at finite temperature.
It would also be interesting to treat the finite-temperature dynamics of
other inverse-square interaction models and compare it to their
zero-temperature exact result \cite{YamamotoJPA32,ArikawaPRL86}.
It comes as a simple generalization of the method used in the present paper.

At the numerical level, finite size calculations can be performed. For 
sufficiently small sizes, the exact correlation functions are accessible using 
the analytical results of this paper. The FFE can not be used, for it is based 
on the thermodynamic limit, rather direct application of (\ref{eq:1point}) is
necessary. Using directly the Yangian multiplets is another solution, as in 
\cite{TalstraPRB50} for zero temperature.

\begin{acknowledgments}
The author thanks K.~Schoutens and J.-S.~Caux for fruitful discussions.
\end{acknowledgments}

\appendix*

\section{Zero-momentum limit of the static structure factor}

In this section we show that
\begin{equation}
F_n = \sum_{m_1>\cdots>m_n} \prod_{i=1}^n n_F(E_{m_i}) \langle
\{m_1,\ldots,m_n\}|p_{-1}p_1|\{m_1,\ldots,m_n\}
\rangle_{\text{irr}}=\int_0^\pi dx \, [n_F(E(x))]^n
\end{equation}
It is clear from the
conservation laws that the intermediate states for each form
factor is only different from the initial state by an increase of
1 of one of the spinons' $m$.
Quite generally, we thus need
\begin{align} |\langle
&\{m_1,\ldots,m_k+1,\ldots,m_n\} | p_1 |
m_1,\ldots,m_k,\ldots,m_n\} \rangle|^2 \simeq \prod_{i \neq k}
\alpha^{\text{sgn}(i-k)}_{(-)^{i-k}} (m_k-m_i) \\
&\alpha^+_\pm (m) =
\left(\frac{\gamma_0(m)\gamma_1(m+1)} {\gamma_1(m)\gamma_0(m+1)} \right)^\pm \\
&\alpha^-_\pm (-m) =\left(\frac{\gamma_1(m)\gamma_0(m-1)}
{\gamma_0(m)\gamma_1(m-1)} \right)^\pm
\end{align}
Explicitly developing the irreducible form factors and separating
different affected $m$'s, we obtain
\begin{align}
F_n &=\sum_{m_1>\cdots>m_n} \prod_{i=1}^n n_F(E_{m_i}) \sum_k
F_n^k
\\ F_n^k &= \sum_{J\subset\{1,\ldots,k-1,k+1,\ldots,n\}} (-)^{n-1-|J|}
\prod_{i\in J} \alpha^{\text{sgn}(i-k)}_{(-)^{d_J(k,i)}}
(m_k-m_i)\\
&=\sum_{J^>\subset\{k+1,\ldots,n\}}(-)^{n-k-|J^>|}\prod_{i\in J^>}
\alpha^+_{(-)^{d_J^>(k,i)}} (m_k-m_i) \,
\sum_{J^<\subset\{1,\ldots,k-1\}} (-)^{k-1-|J^<|}\prod_{i\in J^<}
\alpha^-_{(-)^{d_J^<(k,i)}} (m_i-m_k)
\end{align}
with $d_J(k,i)$ the distance between $k$ and $i$ in the subset
$J$. Thus we can write $F_n^k=F_n^{k>}F_n^{k<}$.

We will know obtain $\sum_{\{m_{i>k}\}} F_n^{k>}$ in a recursive
way (the proof for $F_n^{k>}$ follows the same lines), putting
$k=1$ without loss of generality. We first perform the sum over
$m_n$. Writing $I=\{2,\ldots,n-1\}$, one separates the ensemble of
$J$ as
$$
\{J\subset I \cup \{n\}\} = \{ J + J\cup\{n\}, J\subset I\} = \{ J + J\cup\{n\},
J\in I,|J| \text{ even}\} \cup \{ J + J\cup\{n\}, J\subset I,|J|
\text{ odd}\}.
$$
In the first (resp. second) subset, the distance in J between 1
and $n$ is odd (resp. even). This proves that
\begin{equation}
F_n^{1>} = (\alpha^+_-(m_1-m_n)-1)F_{n-1}^{1>\text{even}} +
(\alpha^+_+(m_1-m_n)-1)F_{n-1}^{1>\text{odd}}
\end{equation}
where the superscripts even and odd corresponds to restrictions of
the expression of $F_{n-1}^{1>}$ to $J$ subsets with even or odd
cardinal. Repeating this recursion one ends up with
\begin{multline}
F_n^{1>} = F_{n-2}^{1>\text{even}}
[(\alpha_+(m_1-m_n)-1)\alpha^+_-(m_1-m_{n-1}) +1-\alpha^+_-(m_1-m_n)]
\\+F_{n-2}^{1>\text{odd}}
[(\alpha^+_-(m_1-m_n)-1)\alpha^+_+(m_1-m_{n-1}) +1-\alpha^+_+(m_1-m_n)]
\end{multline}
Then one easily shows that
\begin{align}
[(\alpha^+_+(m_1-m_n)-1)\alpha^+_-(m_1-m_{n-1})
+1-\alpha^+_-(m_1-m_n)] &\simeq \alpha^+_-(m_1-m_n) \delta_{m_n-m_{n-1}} \\
[(\alpha^+_-(m_1-m_n)-1)\alpha^+_+(m_1-m_{n-1})
+1-\alpha^+_+(m_1-m_n)] &\simeq \alpha^+_+(m_1-m_n)
\delta_{m_n-m_{n-1}}
\end{align}
thus proving the recursion
\begin{equation}
F_n^{1>} = F_{n-1}^{1>}\delta_{m_n-m_{n-1}}
\end{equation}
As this is true only for $n>1$, we obtain for $F_n^k$ the following
\begin{align}
F_n^1 &=(\alpha^+_-(m_1-m_2)-1)\delta_{m_2,\ldots,m_n}\\
F_n^{k\neq 1,n} &= (\alpha^+_-(m_k-m_{k+1})-1)(\alpha^-_+(m_k-m_{k-1})-1)
\delta_{m_1,\ldots,m_{k-1}} \delta_{m_{k+1},\ldots,m_n}\\
F_n^n &=(\alpha^-_+(m_nm_{n-1})-1)\delta_{m_1,\ldots,m_{n-1}}
\end{align}
Now, one has $\alpha^+_-(m_k-m_{k+1})-1 \equiv \delta_{m_k,m_{k+1}}$ and 
$\alpha^-_+(m_k-m_{k-1})-1 \equiv 0$ so that
\begin{equation}
F_n^k = \delta_{k,1} \delta_{m_1,\ldots,m_n}
\end{equation}
Finally
\begin{equation}
F_n=\sum_{m=1}^{N-1} [n_F(E(m))]^n = \int_0^\pi dx \,
[n_F(E(x))]^n
\end{equation}
which ends our proof.

\bibliography{hsbib}

\begin{thebibliography}{39}
\expandafter\ifx\csname natexlab\endcsname\relax\def\natexlab#1{#1}\fi
\expandafter\ifx\csname bibnamefont\endcsname\relax
  \def\bibnamefont#1{#1}\fi
\expandafter\ifx\csname bibfnamefont\endcsname\relax
  \def\bibfnamefont#1{#1}\fi
\expandafter\ifx\csname citenamefont\endcsname\relax
  \def\citenamefont#1{#1}\fi
\expandafter\ifx\csname url\endcsname\relax
  \def\url#1{\texttt{#1}}\fi
\expandafter\ifx\csname urlprefix\endcsname\relax\def\urlprefix{URL }\fi
\providecommand{\bibinfo}[2]{#2}
\providecommand{\eprint}[2][]{\url{#2}}

\bibitem[{\citenamefont{Tennant et~al.}()\citenamefont{Tennant, Lake, Nagler,
  and Frost}}]{TennantEXP}
\bibinfo{author}{\bibfnamefont{D.}~\bibnamefont{Tennant}},
  \bibinfo{author}{\bibfnamefont{B.}~\bibnamefont{Lake}},
  \bibinfo{author}{\bibfnamefont{S.}~\bibnamefont{Nagler}}, \bibnamefont{and}
  \bibinfo{author}{\bibfnamefont{C.}~\bibnamefont{Frost}},
  \bibinfo{note}{unpublished}.

\bibitem[{\citenamefont{Heisenberg}(1928)}]{HeisenbergZP49}
\bibinfo{author}{\bibfnamefont{W.}~\bibnamefont{Heisenberg}},
  \bibinfo{journal}{Zeit. f\"ur Physik} \textbf{\bibinfo{volume}{49}},
  \bibinfo{pages}{619} (\bibinfo{year}{1928}).

\bibitem[{\citenamefont{Bethe}(1931)}]{BetheZP71}
\bibinfo{author}{\bibfnamefont{H.}~\bibnamefont{Bethe}},
  \bibinfo{journal}{Zeit. f\"ur Physik} \textbf{\bibinfo{volume}{71}},
  \bibinfo{pages}{205} (\bibinfo{year}{1931}).

\bibitem[{\citenamefont{M\"uller et~al.}(1981)\citenamefont{M\"uller, Thomas,
  Beck, and Bonner}}]{MuellerPRB24}
\bibinfo{author}{\bibfnamefont{G.}~\bibnamefont{M\"uller}},
  \bibinfo{author}{\bibfnamefont{H.}~\bibnamefont{Thomas}},
  \bibinfo{author}{\bibfnamefont{H.}~\bibnamefont{Beck}}, \bibnamefont{and}
  \bibinfo{author}{\bibfnamefont{J.~C.} \bibnamefont{Bonner}},
  \bibinfo{journal}{Phys. Rev. B} \textbf{\bibinfo{volume}{24}},
  \bibinfo{pages}{1429} (\bibinfo{year}{1981}).

\bibitem[{\citenamefont{Schultz}(1986)}]{SchultzPRB34}
\bibinfo{author}{\bibfnamefont{H.~J.} \bibnamefont{Schultz}},
  \bibinfo{journal}{Phys. Rev. B} \textbf{\bibinfo{volume}{34}},
  \bibinfo{pages}{6372} (\bibinfo{year}{1986}).

\bibitem[{\citenamefont{Jimbo and Miwa}(1995)}]{JimboBOOK}
\bibinfo{author}{\bibfnamefont{M.}~\bibnamefont{Jimbo}} \bibnamefont{and}
  \bibinfo{author}{\bibfnamefont{T.}~\bibnamefont{Miwa}},
  \emph{\bibinfo{title}{Albegraic analysis of solvable lattice models}}
  (\bibinfo{publisher}{American Mathematical Society}, \bibinfo{year}{1995}).

\bibitem[{\citenamefont{Korepin et~al.}(1993)\citenamefont{Korepin, Bogoliubov,
  and Izergin}}]{KorepinBOOK}
\bibinfo{author}{\bibfnamefont{V.~E.} \bibnamefont{Korepin}},
  \bibinfo{author}{\bibfnamefont{N.~M.} \bibnamefont{Bogoliubov}},
  \bibnamefont{and} \bibinfo{author}{\bibfnamefont{A.~G.}
  \bibnamefont{Izergin}}, \emph{\bibinfo{title}{Quantum Inverse Scattering
  Method and Correlation Functions}} (\bibinfo{publisher}{Cambridge Univ.
  Press}, \bibinfo{year}{1993}).

\bibitem[{\citenamefont{Bougourzi et~al.}(1996)\citenamefont{Bougourzi,
  Couture, and Kacir}}]{BougourziPRB54}
\bibinfo{author}{\bibfnamefont{A.}~\bibnamefont{Bougourzi}},
  \bibinfo{author}{\bibfnamefont{M.}~\bibnamefont{Couture}}, \bibnamefont{and}
  \bibinfo{author}{\bibfnamefont{M.}~\bibnamefont{Kacir}},
  \bibinfo{journal}{Phys. Rev. B} \textbf{\bibinfo{volume}{54}},
  \bibinfo{pages}{R12669} (\bibinfo{year}{1996}).

\bibitem[{\citenamefont{Karbach et~al.}(1997)\citenamefont{Karbach, M\"uller,
  Bougourzi, Fledderjohann, and M\"utter}}]{KarbachPRB55}
\bibinfo{author}{\bibfnamefont{M.}~\bibnamefont{Karbach}},
  \bibinfo{author}{\bibfnamefont{G.}~\bibnamefont{M\"uller}},
  \bibinfo{author}{\bibfnamefont{A.}~\bibnamefont{Bougourzi}},
  \bibinfo{author}{\bibfnamefont{A.}~\bibnamefont{Fledderjohann}},
  \bibnamefont{and} \bibinfo{author}{\bibfnamefont{K.-H.}
  \bibnamefont{M\"utter}}, \bibinfo{journal}{Phys. Rev. B}
  \textbf{\bibinfo{volume}{55}}, \bibinfo{pages}{12510} (\bibinfo{year}{1997}).

\bibitem[{\citenamefont{Bougourzi et~al.}(1998)\citenamefont{Bougourzi,
  Karbach, and M\"uller}}]{BougourziPRB57}
\bibinfo{author}{\bibfnamefont{A.}~\bibnamefont{Bougourzi}},
  \bibinfo{author}{\bibfnamefont{M.}~\bibnamefont{Karbach}}, \bibnamefont{and}
  \bibinfo{author}{\bibfnamefont{G.}~\bibnamefont{M\"uller}},
  \bibinfo{journal}{Phys. Rev. B} \textbf{\bibinfo{volume}{57}},
  \bibinfo{pages}{11429} (\bibinfo{year}{1998}).

\bibitem[{\citenamefont{Kitanine et~al.}(2002)\citenamefont{Kitanine, Maillet,
  Slavnov, and Terras}}]{KitanineNPB641}
\bibinfo{author}{\bibfnamefont{N.}~\bibnamefont{Kitanine}},
  \bibinfo{author}{\bibfnamefont{J.-M.} \bibnamefont{Maillet}},
  \bibinfo{author}{\bibfnamefont{N.}~\bibnamefont{Slavnov}}, \bibnamefont{and}
  \bibinfo{author}{\bibfnamefont{V.}~\bibnamefont{Terras}},
  \bibinfo{journal}{Nucl. Phys. B} \textbf{\bibinfo{volume}{641}},
  \bibinfo{pages}{487} (\bibinfo{year}{2002}).

\bibitem[{\citenamefont{Haldane}(1988{\natexlab{a}})}]{HaldanePRL60}
\bibinfo{author}{\bibfnamefont{F.}~\bibnamefont{Haldane}},
  \bibinfo{journal}{Phys. Rev. Lett.} \textbf{\bibinfo{volume}{60}},
  \bibinfo{pages}{635} (\bibinfo{year}{1988}{\natexlab{a}}).

\bibitem[{\citenamefont{Shastry}(1988)}]{ShastryPRL60}
\bibinfo{author}{\bibfnamefont{B.}~\bibnamefont{Shastry}},
  \bibinfo{journal}{Phys. Rev. Lett.} \textbf{\bibinfo{volume}{60}},
  \bibinfo{pages}{639} (\bibinfo{year}{1988}).

\bibitem[{\citenamefont{Haldane}(1988{\natexlab{b}})}]{HaldanePRL66}
\bibinfo{author}{\bibfnamefont{F.}~\bibnamefont{Haldane}},
  \bibinfo{journal}{Phys. Rev. Lett.} \textbf{\bibinfo{volume}{66}},
  \bibinfo{pages}{1529} (\bibinfo{year}{1988}{\natexlab{b}}).

\bibitem[{\citenamefont{Haldane}(1991)}]{HaldanePRL67}
\bibinfo{author}{\bibfnamefont{F.}~\bibnamefont{Haldane}},
  \bibinfo{journal}{Phys. Rev. Lett.} \textbf{\bibinfo{volume}{67}},
  \bibinfo{pages}{937} (\bibinfo{year}{1991}).

\bibitem[{\citenamefont{Haldane and Zirnbauer}(1993)}]{HaldanePRL71}
\bibinfo{author}{\bibfnamefont{F.}~\bibnamefont{Haldane}} \bibnamefont{and}
  \bibinfo{author}{\bibfnamefont{M.}~\bibnamefont{Zirnbauer}},
  \bibinfo{journal}{Phys. Rev. Lett.} \textbf{\bibinfo{volume}{71}},
  \bibinfo{pages}{4055} (\bibinfo{year}{1993}).

\bibitem[{\citenamefont{Uglov}(1998)}]{UglovCMP191}
\bibinfo{author}{\bibfnamefont{D.}~\bibnamefont{Uglov}},
  \bibinfo{journal}{Commun. Math. Phys.} \textbf{\bibinfo{volume}{191}},
  \bibinfo{pages}{663} (\bibinfo{year}{1998}).

\bibitem[{\citenamefont{Yamamoto
  et~al.}(2000{\natexlab{a}})\citenamefont{Yamamoto, Saiga, Arikawa, and
  Kuramoto}}]{YamamotoJPSJ69}
\bibinfo{author}{\bibfnamefont{T.}~\bibnamefont{Yamamoto}},
  \bibinfo{author}{\bibfnamefont{Y.}~\bibnamefont{Saiga}},
  \bibinfo{author}{\bibfnamefont{M.}~\bibnamefont{Arikawa}}, \bibnamefont{and}
  \bibinfo{author}{\bibfnamefont{Y.}~\bibnamefont{Kuramoto}},
  \bibinfo{journal}{J. Phys. Soc. Jap.} \textbf{\bibinfo{volume}{69}},
  \bibinfo{pages}{900} (\bibinfo{year}{2000}{\natexlab{a}}).

\bibitem[{\citenamefont{Yamamoto
  et~al.}(2000{\natexlab{b}})\citenamefont{Yamamoto, Saiga, Arikawa, and
  Kuramoto}}]{YamamotoPRL84}
\bibinfo{author}{\bibfnamefont{T.}~\bibnamefont{Yamamoto}},
  \bibinfo{author}{\bibfnamefont{Y.}~\bibnamefont{Saiga}},
  \bibinfo{author}{\bibfnamefont{M.}~\bibnamefont{Arikawa}}, \bibnamefont{and}
  \bibinfo{author}{\bibfnamefont{Y.}~\bibnamefont{Kuramoto}},
  \bibinfo{journal}{Phys. Rev. Lett.} \textbf{\bibinfo{volume}{84}},
  \bibinfo{pages}{1308} (\bibinfo{year}{2000}{\natexlab{b}}).

\bibitem[{\citenamefont{Yamamoto and Arikawa}(1999)}]{YamamotoJPA32}
\bibinfo{author}{\bibfnamefont{T.}~\bibnamefont{Yamamoto}} \bibnamefont{and}
  \bibinfo{author}{\bibfnamefont{M.}~\bibnamefont{Arikawa}},
  \bibinfo{journal}{J. Phys. A: Math. Gen.} \textbf{\bibinfo{volume}{32}},
  \bibinfo{pages}{3341} (\bibinfo{year}{1999}).

\bibitem[{\citenamefont{Arikawa et~al.}(1999)\citenamefont{Arikawa, Yamamoto,
  Saiga, and Kuramoto}}]{ArikawaJPSJ68}
\bibinfo{author}{\bibfnamefont{M.}~\bibnamefont{Arikawa}},
  \bibinfo{author}{\bibfnamefont{T.}~\bibnamefont{Yamamoto}},
  \bibinfo{author}{\bibfnamefont{Y.}~\bibnamefont{Saiga}}, \bibnamefont{and}
  \bibinfo{author}{\bibfnamefont{Y.}~\bibnamefont{Kuramoto}},
  \bibinfo{journal}{J. Phys. Soc. Jap.} \textbf{\bibinfo{volume}{68}},
  \bibinfo{pages}{3782} (\bibinfo{year}{1999}).

\bibitem[{\citenamefont{Arikawa et~al.}(2001)\citenamefont{Arikawa, Saiga, and
  Kuramoto}}]{ArikawaPRL86}
\bibinfo{author}{\bibfnamefont{M.}~\bibnamefont{Arikawa}},
  \bibinfo{author}{\bibfnamefont{Y.}~\bibnamefont{Saiga}}, \bibnamefont{and}
  \bibinfo{author}{\bibfnamefont{Y.}~\bibnamefont{Kuramoto}},
  \bibinfo{journal}{Phys. Rev. Lett.} \textbf{\bibinfo{volume}{86}},
  \bibinfo{pages}{3096} (\bibinfo{year}{2001}).

\bibitem[{\citenamefont{LeClair and Mussardo}(1999)}]{LeClairNPB552}
\bibinfo{author}{\bibfnamefont{A.}~\bibnamefont{LeClair}} \bibnamefont{and}
  \bibinfo{author}{\bibfnamefont{G.}~\bibnamefont{Mussardo}},
  \bibinfo{journal}{Nucl. Phys. B} \textbf{\bibinfo{volume}{552}},
  \bibinfo{pages}{624} (\bibinfo{year}{1999}).

\bibitem[{\citenamefont{Delfino}(2001)}]{DelfinoJPA34}
\bibinfo{author}{\bibfnamefont{G.}~\bibnamefont{Delfino}}, \bibinfo{journal}{J.
  Phys. A: Math. Gen.} \textbf{\bibinfo{volume}{34}}, \bibinfo{pages}{L161}
  (\bibinfo{year}{2001}).

\bibitem[{\citenamefont{Mussardo}(2001)}]{MussardoJPA34}
\bibinfo{author}{\bibfnamefont{G.}~\bibnamefont{Mussardo}},
  \bibinfo{journal}{J. Phys. A: Math. Gen.} \textbf{\bibinfo{volume}{34}},
  \bibinfo{pages}{7399} (\bibinfo{year}{2001}).

\bibitem[{\citenamefont{Konik}()}]{KonikXXX}
\bibinfo{author}{\bibfnamefont{R.}~\bibnamefont{Konik}},
  \bibinfo{note}{arXiv:cond-mat/0105284}.

\bibitem[{\citenamefont{Saleur}(2000)}]{SaleurNPB567}
\bibinfo{author}{\bibfnamefont{H.}~\bibnamefont{Saleur}},
  \bibinfo{journal}{Nucl. Phys. B} \textbf{\bibinfo{volume}{567}},
  \bibinfo{pages}{602} (\bibinfo{year}{2000}).

\bibitem[{\citenamefont{Castro-Alvaredo and
  Fring}(2002)}]{Castro-AlvaredoNPB636}
\bibinfo{author}{\bibfnamefont{O.}~\bibnamefont{Castro-Alvaredo}}
  \bibnamefont{and} \bibinfo{author}{\bibfnamefont{A.}~\bibnamefont{Fring}},
  \bibinfo{journal}{Nucl. Phys. B} \textbf{\bibinfo{volume}{636}},
  \bibinfo{pages}{611} (\bibinfo{year}{2002}).

\bibitem[{\citenamefont{van Elburg and Schoutens}(2000)}]{vanElburgJPA33}
\bibinfo{author}{\bibfnamefont{R.}~\bibnamefont{van Elburg}} \bibnamefont{and}
  \bibinfo{author}{\bibfnamefont{K.}~\bibnamefont{Schoutens}},
  \bibinfo{journal}{J. Phys. A: Math. Gen.} \textbf{\bibinfo{volume}{33}},
  \bibinfo{pages}{7987} (\bibinfo{year}{2000}).

\bibitem[{\citenamefont{Peysson and Schoutens}(2002)}]{PeyssonJPA35}
\bibinfo{author}{\bibfnamefont{S.}~\bibnamefont{Peysson}} \bibnamefont{and}
  \bibinfo{author}{\bibfnamefont{K.}~\bibnamefont{Schoutens}},
  \bibinfo{journal}{J. Phys. A: Math. Gen.} \textbf{\bibinfo{volume}{35}},
  \bibinfo{pages}{6471} (\bibinfo{year}{2002}).

\bibitem[{\citenamefont{Polychronakos}(1993)}]{PolychronakosPRL70}
\bibinfo{author}{\bibfnamefont{A.}~\bibnamefont{Polychronakos}},
  \bibinfo{journal}{Phys. Rev. Lett.} \textbf{\bibinfo{volume}{70}},
  \bibinfo{pages}{2329} (\bibinfo{year}{1993}).

\bibitem[{\citenamefont{Ha and Haldane}(1992)}]{HaPRB46}
\bibinfo{author}{\bibfnamefont{Z.}~\bibnamefont{Ha}} \bibnamefont{and}
  \bibinfo{author}{\bibfnamefont{F.}~\bibnamefont{Haldane}},
  \bibinfo{journal}{Phys. Rev. B} \textbf{\bibinfo{volume}{46}},
  \bibinfo{pages}{9359} (\bibinfo{year}{1992}).

\bibitem[{\citenamefont{Macdonald}(1995)}]{Macdonald}
\bibinfo{author}{\bibfnamefont{I.}~\bibnamefont{Macdonald}},
  \emph{\bibinfo{title}{Symmetric functions and Hall polynomials}}
  (\bibinfo{publisher}{Oxford Univ. Press}, \bibinfo{year}{1995}).

\bibitem[{\citenamefont{Bouwknegt and Schoutens}(1999)}]{BouwknegtNPB547}
\bibinfo{author}{\bibfnamefont{P.}~\bibnamefont{Bouwknegt}} \bibnamefont{and}
  \bibinfo{author}{\bibfnamefont{K.}~\bibnamefont{Schoutens}},
  \bibinfo{journal}{Nucl. Phys. B} \textbf{\bibinfo{volume}{547}},
  \bibinfo{pages}{501} (\bibinfo{year}{1999}).

\bibitem[{\citenamefont{Isakov}(1994)}]{IsakovMPLB8}
\bibinfo{author}{\bibfnamefont{S.}~\bibnamefont{Isakov}},
  \bibinfo{journal}{Mod. Phys. Lett. B} \textbf{\bibinfo{volume}{8}},
  \bibinfo{pages}{319} (\bibinfo{year}{1994}).

\bibitem[{\citenamefont{Dasni\`eres~de Veigy and
  Ouvry}(1994)}]{DasnieresdeVeigyPRL72}
\bibinfo{author}{\bibfnamefont{A.}~\bibnamefont{Dasni\`eres~de Veigy}}
  \bibnamefont{and} \bibinfo{author}{\bibfnamefont{S.}~\bibnamefont{Ouvry}},
  \bibinfo{journal}{Phys. Rev. Lett.} \textbf{\bibinfo{volume}{72}},
  \bibinfo{pages}{600} (\bibinfo{year}{1994}).

\bibitem[{\citenamefont{Wu}(1994)}]{WuPRL73}
\bibinfo{author}{\bibfnamefont{Y.}~\bibnamefont{Wu}}, \bibinfo{journal}{Phys.
  Rev. Lett.} \textbf{\bibinfo{volume}{73}}, \bibinfo{pages}{922}
  (\bibinfo{year}{1994}).

\bibitem[{\citenamefont{LeClair et~al.}(1996)\citenamefont{LeClair, Lesage,
  Sachdev, and Saleur}}]{LeClairNPB482}
\bibinfo{author}{\bibfnamefont{A.}~\bibnamefont{LeClair}},
  \bibinfo{author}{\bibfnamefont{F.}~\bibnamefont{Lesage}},
  \bibinfo{author}{\bibfnamefont{S.}~\bibnamefont{Sachdev}}, \bibnamefont{and}
  \bibinfo{author}{\bibfnamefont{H.}~\bibnamefont{Saleur}},
  \bibinfo{journal}{Nucl. Phys. B} \textbf{\bibinfo{volume}{482}},
  \bibinfo{pages}{579} (\bibinfo{year}{1996}).

\bibitem[{\citenamefont{Talstra and Haldane}(1994)}]{TalstraPRB50}
\bibinfo{author}{\bibfnamefont{J.}~\bibnamefont{Talstra}} \bibnamefont{and}
  \bibinfo{author}{\bibfnamefont{F.}~\bibnamefont{Haldane}},
  \bibinfo{journal}{Phys. Rev. B} \textbf{\bibinfo{volume}{50}},
  \bibinfo{pages}{6889} (\bibinfo{year}{1994}).

\end{thebibliography}

\end{document}